\documentclass{aa}  

\usepackage{graphicx}
\usepackage{txfonts}
\newcommand\un[1]{{\,\rm #1}}
\newcommand\E[1]{\times10^{#1}}
\newcommand\rs[1]{_\mathrm{#1}}
\newcommand\g{$\gamma$}

\begin{document} 

   \title{GeV light curves of young supernova remnants}

   \author{O. Petruk\inst{1,2},
          V. Beshley\inst{1},
          V. Marchenko\inst{3},
          M. Patrii\inst{4}
          }

   \institute{Institute for Applied Problems in Mechanics and Mathematics, Naukova 3-b, 79060 Lviv, Ukraine\\
              \email{beshley.vasyl@gmail.com}
        \and
        Astronomical Observatory, Ivan Franko National University of Lviv, Kyryla and Methodia St. 8, 79005 Lviv, Ukraine
        \and
        Astronomical Observatory, Jagiellonian University, Orla 171, 30-244 Cracow, Poland
        \and
        Faculty of Physics, Ivan Franko National University of Lviv, Kyryla and Methodia St. 8, 79005 Lviv, Ukraine\\
             }

   \date{Received ... .., 2019; accepted ... ..., 2019}
  \abstract
  {Observational data from the Fermi Gamma-ray Space Telescope are analyzed with a goal in mind to look for variations in \g-ray flux from young shell-like supernova remnants. Uniform methodological approach is adopted for all SNRs considered.   G1.9+0.3 and Kepler SNRs are not detected. The light curves of Cas~A and Tycho SNRs are compatible with the steady GeV flux during the recent ten years, as also X-ray and radio fluxes. Less confident results on SN1006 and SN1987A are discussed. 
  }


   \maketitle
%

\section{Introduction}

Almost two decades passed after the first view of a shell-like supernova remnant (SNR) in \g-rays:  HEGRA stereoscopic system has detected a flux from Cas~A  \citep{2001A&A...370..112A}. Next generation of Cherenkov telescopes as well as the Fermi observatory advance our knowledge about \g-ray emission from SNRs. At present, there are 13 firm confirmations of TeV gamma-rays from the shell-like galactic SNRs: 10 observed by H.E.S.S. \citep{2018A&A...612A...3H,2018A&A...612A...8H} and 3 by northern Cherenkov observatories \citep{2007A&A...474..937A,2007ApJ...664L..87A,2011ApJ...730L..20A}. In addition to this, 8 H.E.S.S. sources are composite SNRs (consisting of pulsar and SNR shell), and almost 20 H.E.S.S. sources are SNR candidates \citep{2018A&A...612A...3H,2018A&A...612A...8H}. Systematic search of SNRs in a softer GeV photon energy range are presented in the Fermi SNR catalogue \citep{2016ApJS..224....8A}.\footnote{\url{https://fermi.gsfc.nasa.gov/ssc/data/access/lat/1st_SNR_catalog/}} There are 30 SNRs and 14 possible SNRs have been listed in this reference. 
A useful tool for those interested in the high-energy emission from SNRs is a catalogue first introduced by \citet{2012AdSpR..49.1313F}.\footnote{\url{http://snrcat.physics.umanitoba.ca/}} 

All \g-ray observations of SNRs are a clear demonstration that SNR shocks are able to accelerate cosmic-rays (CRs) to multi-TeV energies. The shapes of spectra give deeper inside into related physics. In particular, detection of the high-energy cut-off in the \g-ray spectrum of Cas~A \citep{2017MNRAS.472.2956A} and other SNRs is somehow pity: it demonstrates that at least these SNRs do not accelerate cosmic rays to the energies of the knee in the CR spectrum. The low-energy portions of \g-ray spectra in IC443 and W44 appear to be more impressive: they are the first observational proofs the protons accelerated in the SNR shocks \citep{2013Sci...339..807A}. 

In contrast to SNRs, no supernova has been observed in \g-rays (except of the long GRBs which are believed to arise from the core-collapse event). Therefore, there is also no observational clue of how \g-ray light curves of these explosive events look like. 

There is a sign of the temporal evolution of the \g-ray spectra from SNRs -- on a time-scale of the SNR lifespan: spectra of the young, middle-age and old SNRs seem to group separately \citep[Fig.~6 in][]{2015ARNPS..65..245F}. Such a property reflects mostly the cardinal changes in SNRs properties relevant for different evolutionary stages.

Time evolution of \g-ray emission from shells of young SNRs could be important as an insight to physics of the time-dependent particle acceleration at the fresh shocks. To this reason, the remnant of SN1987A would be a promising source. An attractive result related to the possible detection of GeV \g-rays from SN1987 has been published recently by \citet{2019arXiv190303045M}. Other historical SNRs are interesting in this respect as well. 

The goal of the present paper is to look for the time variations in the GeV \g-ray fluxes from young SNRs. 
Namely, we analyse, under the uniform approach, the \g-rays from the historical supernova remnants, with the age up to a thousand years: SN1987A, G1.9+0.3, Cas~A, Kepler, Tycho, SN1006. We primarily have in mind the shock particle acceleration, therefore, we are interested in the shell-like SNRs only. Therefore, the pulsar-dominated historical SNRs, Crab nebula and 3C58, are not considered in the present paper.

\begin{table*}
\caption{GeV fluxes of the historical shell-like SNRs. Comparison with previous measurements. In cases of G1.9+0.3 and Kepler, the upper limits are presented.}
\centering
\begin{tabular}{llllllll}
\hline\hline
     & SN      &     & energy & observational & flux or      & flux or         & \\
SNR  & event   & Ref & range  & data from     & upper limit  &  upper limit    & units\\
     & year    &     & GeV    & the period    & (reference)  & (present paper) & \\

\hline
SN1987A & 1987& 3& 1.0--100& {08.2016--12.2018}& {$8.3\pm2.4$}& $9.7\pm2.7$& $10^{-10}\un{ph\ cm^{-2}s^{-1}}$ \\
G1.9+0.3 & 1900$^{(1)}$& 4&0.2--300 &08.2008--06.2014 & 4.43& 0.22 & $10^{-9}\un{ph\ cm^{-2}s^{-1}}$ \\
Cas~A & 1680$^{(2)}$& 5& 0.1--100& 08.2008--04.2012& $6.2\pm0.4$& $5.8\pm0.3$& $10^{-11}\un{erg\ cm^{-2}s^{-1}}$ \\
Kepler & 1604& -- &0.1--100 &08.2009--08.2019 & -- & 2.73 &$10^{-10}\un{ph\ cm^{-2}s^{-1}}$ \\
Tycho & 1572& 6& 0.4--100& 08.2008--05.2011& $3.5\pm1.1$& $2.6\pm0.5$& $10^{-9}\un{ph\ cm^{-2}s^{-1}}$ \\
SN1006 & 1006& 7& 0.5--500& 08.2008--12.2018& $5.9\pm1.7$& $5.8\pm1.2$& $10^{-10}\un{ph\ cm^{-2}s^{-1}}$ \\
\hline
\end{tabular}
\tablebib{(1)~\citet{2011ApJ...737L..22C}; (2)~\citet{1980JHA....11....1A}; (3)~\citet{2019arXiv190303045M}; (4)~\citet{2015AdSpR..56.1793G}; (5)~\citet{2013ApJ...779..117Y}; (6)~\citet{2012ApJ...744L...2G};
(7)~\citet{2019PASJ...71...77X}}
\label{gev:table-i}
\end{table*}

\section{Data analysis}
\label{gev:sectdataan}

The standard binned likelihood analysis with \textit{gtlike}\footnote{\url{https://fermi.gsfc.nasa.gov/ssc/data/analysis/}} tool provided by Fermi Science Tools are used for all SNRs presented in this paper. We use the latest release of the LAT Pass 8 data and consider period that covers 10 years of observation (from August 2009 to August 2019). The regions of interest (ROI) is centered on a given object and has radius $14^\circ$. All events with a zenith angle greater than $90^\circ$ (see LAT instrument team recommendation) are rejected. The last version of the Fermi Science Tool with the \verb+P8R3_CLEAN_V2+  Instrumental Response Function is used. Upper limits are calculated with \textit{UpperLimits} python module. 

The method takes into account all known neighbour and background sources which are in ROI. The model includes all sources from the 4FGL catalogue (\verb+gll_psc_v19.fit+) as well as both the Galactic diffuse background \verb+gll_iem_v07.fits+ and extra-galactic isotropic background \verb+iso_P8R3_CLEAN_V2_v1.txt+.
The spectra of the neighbour sources are the same as in the 4FGL catalogue. The normalisation parameter are free for all objects in ROI; other parameters are fixed. We also include the sources with fixed parameters which are within an annulus with size from $14^\circ$ to $28^\circ$ outside of ROI region. All sources which appears to have $TS<0$ were excluded from consideration.

In the present paper we analyse six young SNRs. Three of them are listed in the 4FGL catalogue \citep{2019arXiv190210045T}\footnote{\url{https://fermi.gsfc.nasa.gov/ssc/data/access/lat/8yr_catalog/}}, namely, Cas~A (4FGL J2323.4+5849), Tycho (4FGL J0025.3+6408) and SN1006 (4FGL J1503.6-4146). 
Therefore, we used for each of these SNRs the model with \verb+PLSuperExpCutoff2+ spectrum\footnote{\url{https://fermi.gsfc.nasa.gov/ssc/data/analysis/scitools/source_models.html}} with a free normalization. Aa to the other three SNRs, SN1987a, G1.9+0.3 and Kepler, we adopt a model with the power-low spectrum with free normalization and the spectral index 2.1 \citep{2019arXiv190303045M}, 2.6 \citep{2015AdSpR..56.1793G}, 2.0 (lower index in \citet{2008A&A...488..219A}) respectively.

\begin{figure}
\resizebox{\hsize}{!}{\includegraphics{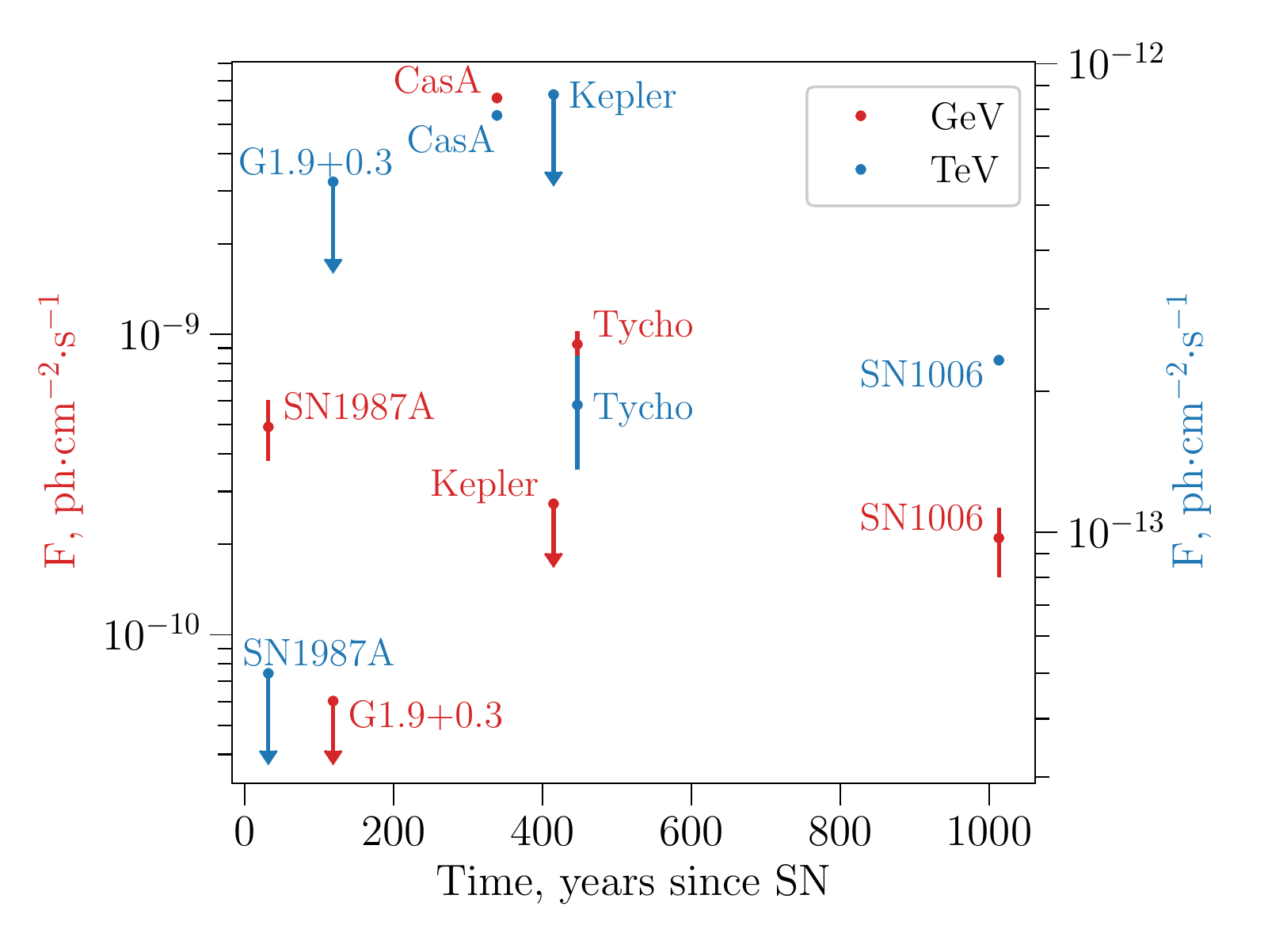}}
\caption{Flux versus age for young historical shell-like SNRs. Red data points (the left axis) represent the GeV fluxes in the photon energies 1-100 GeV and for 10 years of observations. Values of the test statistics of our analysis for the red points: SN1987A ($TS=36$), G1.9+0.3 ($TS\approx0$), CasA ($TS=3642$), Kepler ($TS=4.9$), Tycho ($TS=132$), SN1006 ($TS=23$). Errors are at the $1\sigma$ level. 
Blue data points (the right axis) show the TeV fluxes (for photon energies $>1$ TeV). References for the blue points: SN1987A \citep{2015Sci...347..406H}; G1.9+0.3 \citep{2014MNRAS.441..790H}; Cas~A \citep{2010ApJ...714..163A}; Kepler \citep{2008A&A...488..219A}; Tycho \citep{2011ApJ...730L..20A}; SN1006 \citep{2010A&A...516A..62A}.}
\label{gev:fagetev}
\end{figure}

\section{Results}

\subsection{GeV fluxes}

First, we checked if our methodology recovers known measurements. 
Table \ref{gev:table-i} presents the list of the historical shell-like SNRs with the \g-ray fluxes from the literature (sixth column). Fluxes calculated with our approach (in the seventh column) correspond to the same observational period and the same photon energy range as in the reference. The differences are within the errors and appear because we used the last data release of (P8R3) and the most recent source and background models.

Then, we performed a uniform analysis of the Fermi data for all these SNRs. At the beginning, we have measured the fluxes in the photon energy range $1-100\un{GeV}$ for 10 years of observations (August 2009 -- August 2019) and plotted them versus the SNR age (Fig.~\ref{gev:fagetev} red points). GeV fluxes are shown on this figure for the objects with the test statistics $TS>20$; the  upper limits are shown for the smaller $TS$. 

The GeV signal from Kepler SNR has $TS=4.9$, therefore, one cannot say about its detection in the data collected by Fermi LAT during 10 years of observation. The upper limit for the total photon flux is $2.7\E{-10}\un{ph\ cm^{-2}\ s^{-1}}$. G1.9+0.3 is not detected as well ($TS$ is close to zero) with the upper limit $6.0\E{-11}\un{ph\ cm^{-2}\ s^{-1}}$. 
Kepler SNR is within ROI for G1.9+0.3 (distance from the center is about $7^\circ$). Its test statistics is larger than zero, therefore, we included Kepler SNR in the model for G1.9+0.3, with the power-low spectrum with free normalization and the spectral index given at the end of Sect.~\ref{gev:sectdataan}. 

Looking at Fig.~\ref{gev:fagetev}, we cannot infer any meaningful hint about an evolution of the GeV \g-ray flux in young SNRs. Different SN types and expansion in different ambient conditions are rather more important in determination of the individual SNR flux than eventual similarities in the early evolution of the particle acceleration. 

TeV fluxes look to behave in a similar way to the GeV fluxes, as it may be seen on Fig.~\ref{gev:fagetev} for Cas~A, Tycho and SN1006. This is not surprise because GeV and TeV fluxes are parts of the broader spectra of the same SNRs. Interestingly that TeV fluxes for Cas~A and Tycho are lower than the GeV fluxes, in contrast to SN1006 where the \g-ray spectrum is harder.

\begin{figure}
\resizebox{\hsize}{!}{\includegraphics[trim={14pt 10pt 14pt 17pt},clip]{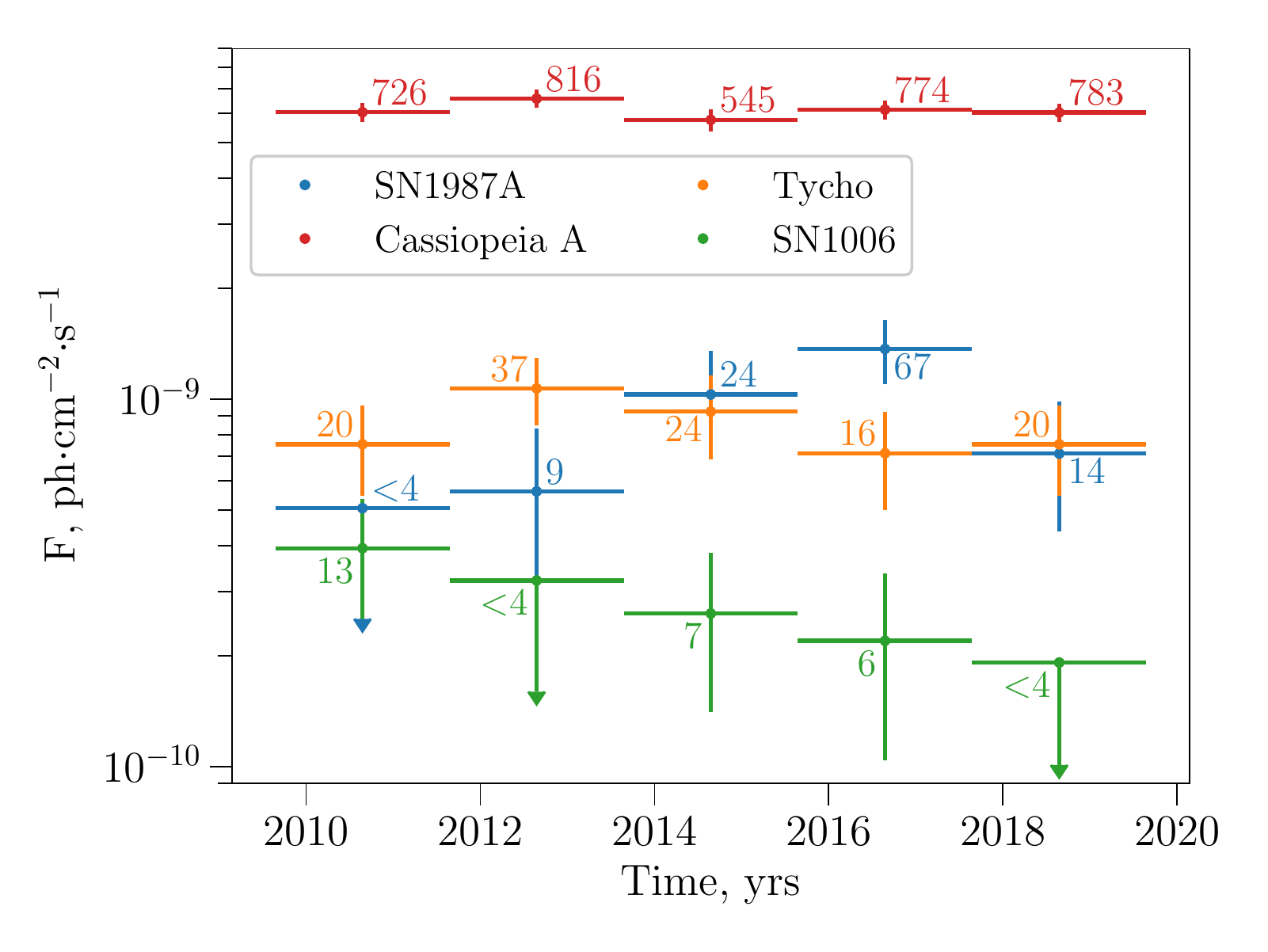}}
\caption{Flux evolution for four supernova remnants in the photon energy range 1-100~GeV. One point corresponds to 2 years of Fermi LAT observations. The values of TS are shown near the data points.}
\label{gev:lightcurves}
\end{figure}

\subsection{Light curves}

Four young SNRs have rather high detection significance in the Fermi LAT data for 10 years of observations. Therefore, we have analysed fluxes of these SNRs in the consecutive 2-year time intervals. Results are shown on Fig.~\ref{gev:lightcurves}. For this plot, the upper limits were calculated if the test statistic for a given point is less than $4$. 

Such a plot is more informative in respect of the temporal variation in the \g-ray emission of young SNRs. Each point in the light curves for Cas~A and Tycho has significance above $4\sigma$. Four of five points for SN1987A are above $3\sigma$. Trend for SN1006 is less significant. 

Light curves for Cas~A and Tycho in \g-rays are compatible with constant emission over the recent 10 years. 

\section{Discussion}

\begin{table}
\centering
\caption{Observation log for the analyzed Chandra data}
\label{gev:tab:Chandra_data}
\begin{tabular}{llc}
\hline
\multicolumn{1}{c}{ObsID} & \multicolumn{1}{c}{Start date [UT]} & \multicolumn{1}{c}{Exposure [ks]} \\
\hline
\multicolumn{3}{c}{Cas~A} \\
\hline
4638   &    2004-04-14 19:47:55	 & 164.53	 \\
9117   &    2007-12-05 22:00:54	 & 24.84	 \\
10935  &    2009-11-02 22:16:52	 & 23.26	 \\
14229  &    2012-05-15 09:15:11	 & 49.09 \\
18344  &    2016-10-21 16:58:44	 & 25.75 \\
19605  &    2018-05-15 16:06:34	 & 49.41 \\
\hline
\multicolumn{3}{c}{Tycho}   \\
\hline
3837  &   2003-04-29 01:59:47	 & 145.6 \\
7639  &   2007-04-23 02:18:40	 & 108.87 \\
10095 &   2009-04-23 21:27:53	 & 173.37 \\
15998 &   2015-04-22 22:19:05	 & 146.98 \\
\hline
\multicolumn{3}{c}{SN1006} \\
\hline
9107  &  2003-04-08 06:33:03	   & 20.13 \\
3838  &  2008-06-24 14:04:29	   & 68.87 \\
\hline 
\hline
\end{tabular}
\end{table}

\begin{figure}
\resizebox{\hsize}{!}{\includegraphics[trim={0 -4pt 0 0},clip]{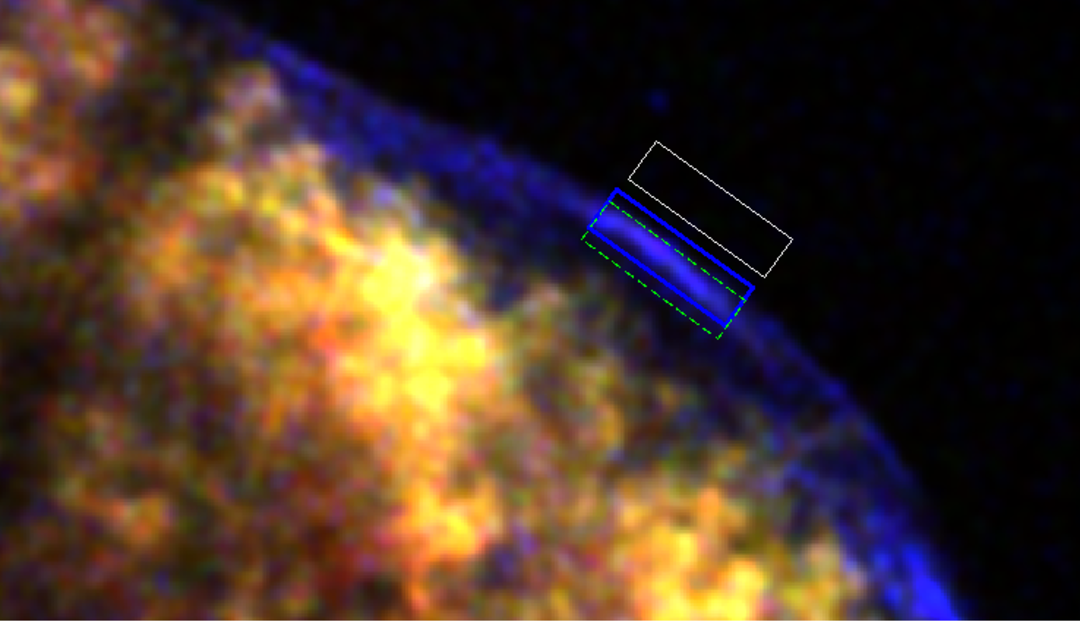}}
\resizebox{\hsize}{!}{\includegraphics[trim={0 -4pt 0 0},clip]{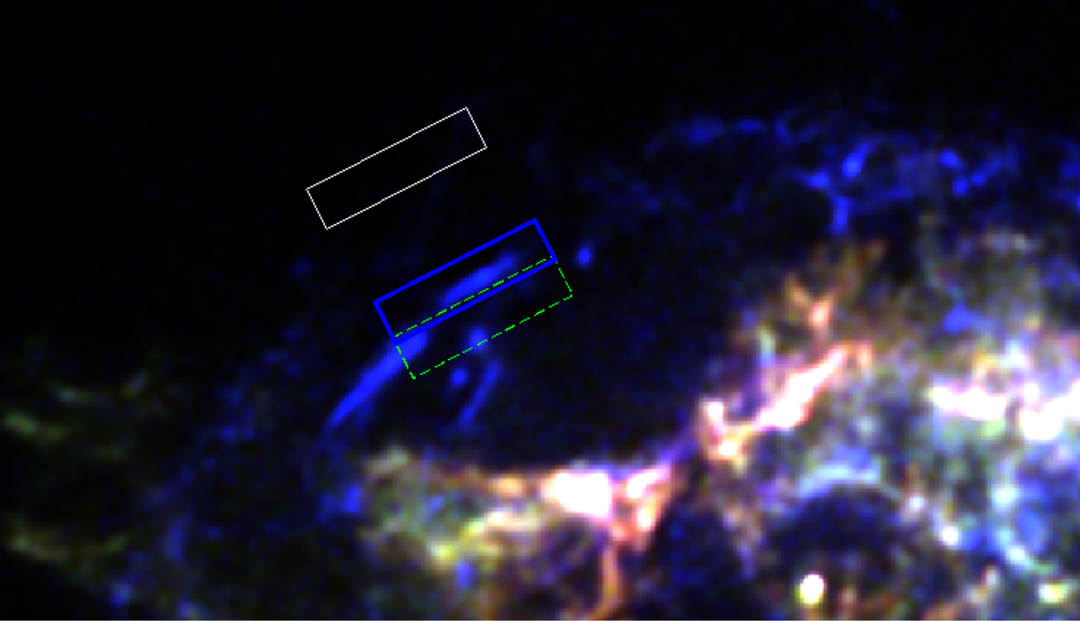}}
\resizebox{\hsize}{!}{\includegraphics{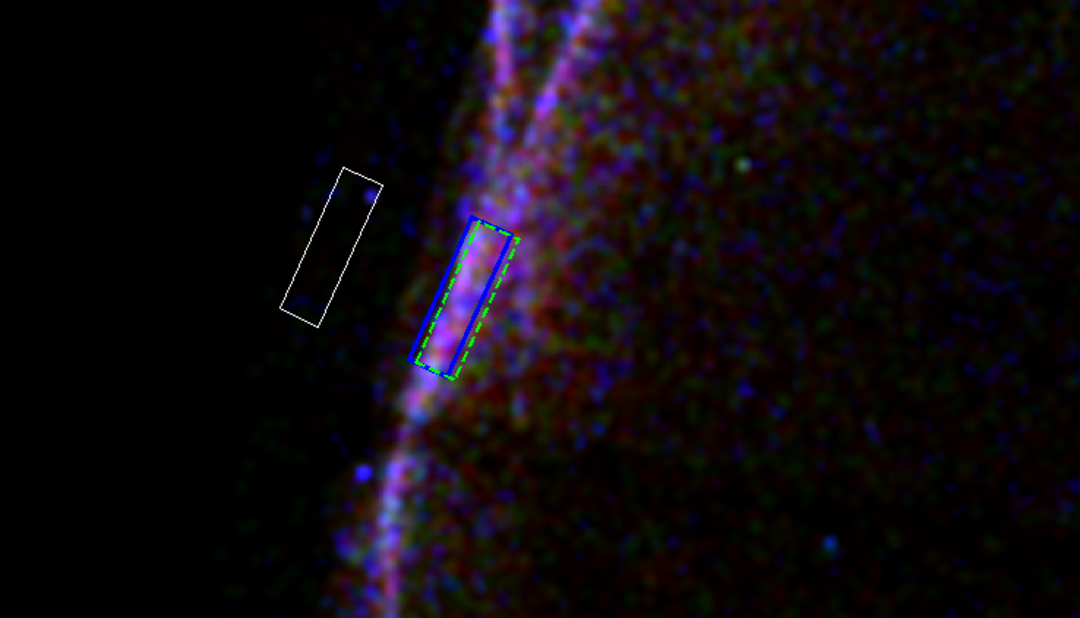}}
\caption{Images of region around the forward shock in SNRs: Tycho (top, obsID 15998, 22 April 2015), Cas~A (middle, obsID 19605, 15 May 2018), SN1006 (bottom, obsID 9107, 24 June 2008). Colors: 1.7-2.1 keV (red, Si line), 2.3-2.6 keV (green, S line), 4.0-6.0 keV (blue, non-thermal). Blue rectangle marks the region where the flux is measured and the white one shows the region for the background emission. Their locations corresponds to the time of observation presented. Green rectangle corresponds to the location of the same shock region in 2003 (Tycho), 2004 (Cas~A), 2000 (SN1006) year. Differences between the green and blue rectangles are due to SNR expansion.}
\label{gev:xrays}
\end{figure}

\subsection{Comparison with X-rays}

Are the \g-ray flux variations for these SNRs similar to the evolution of the non-thermal emission in the X-ray band? 

In order to answer this question, we have analysed X-ray emission from small regions containing the forward shock where the non-thermal emission dominates the thermal one. 
We have selected several Chandra observations of three SNRs performed since 2003 (see Table ~\ref{gev:tab:Chandra_data} for details). After choosing a region, all the observations containing this region were processed; a region was shifted between different observation ID in accordance to the SNR expansion in order to ensure we measure the flux variation from the same portion of the shock. 

The analysis was carried out using the software package {\tt CIAO\,4.11} \citep{2007ChNew..14...36F} and the calibration database {\tt CALDB\,4.8.3}. Before the analysis, the data were reprocessed using the {\tt chandra\_repro} script, following the standard recommendations of the {\tt CIAO} analysis threads. 

For the spectral analysis we used the {\tt Sherpa} fitting application \citep{2001SPIE.4477...76F}. After the extraction the source and background spectra, the background spectra was subtracted and the fitting was performed for the grouped spectra with the minimum signal-to-noise ratio 3. 

For spatial analysis the data were merged and binned with a binning factor of 1, which corresponds to the original {\it Chandra} pixel size of $0.492\arcsec$. 

For the estimation of flux in source regions of Chandra data with corresponding background regions we used {\tt srcflux} tool from {\tt CIAO} software. The flux was calculated using a model dependent estimate, where as a model we used simple absorbed power law 
{\tt xsphabs*xspowerlaw}.

In order to visualize the differences in spatial structures of the regions around the forward shock in SNRs in different X-ray energy bands, we have used the three-color capabilities of {\tt SAOImageDS9} program, where data in the three different energy bands were loaded into RGB frame and highlighted in appropriate colors (Fig.~\ref{gev:xrays}). The images of the selected observations were smoothed a bit with 2D Gaussian to improve visual appearance. Actually, the images in colour were used to choose a region around the forward shock where the hard X-ray emission dominates.

\begin{figure}
\resizebox{\hsize}{!}{\includegraphics[trim={14pt 10pt 14pt 17pt},clip]{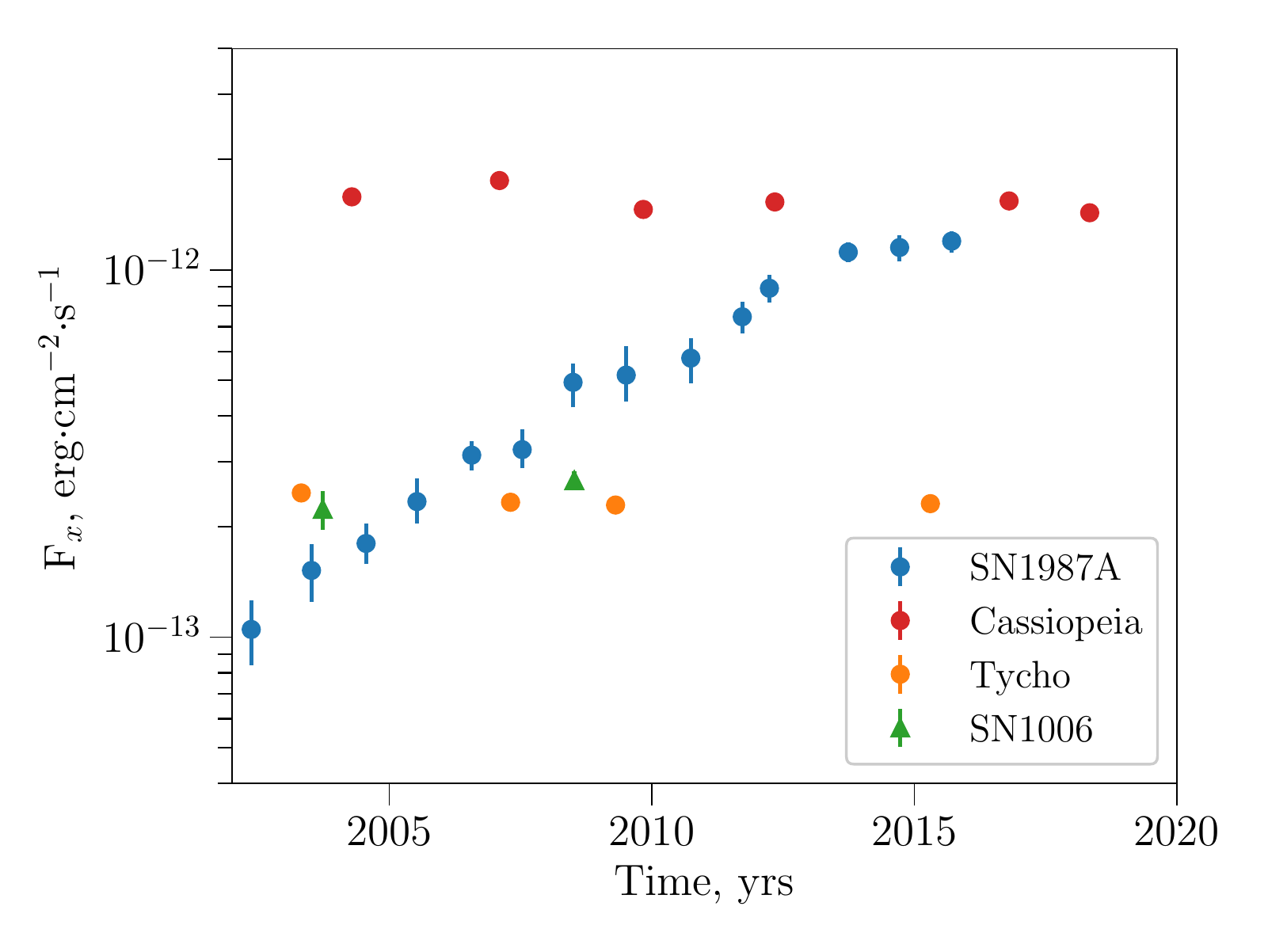}}
\caption{Evolution of the X-ray fluxes from the shock regions (shown on Fig.~\ref{gev:xrays}) for Cas~A, Tycho, SN1006 in the photon energy range 3-7 keV. Errors represent $90\%$ confidence interval; they are smaller than the dot sizes in some cases. Fluxes for SN1987A are from the whole SNR and for the photon energies 3-8 keV \citep{2016ApJ...829...40F}. Colors are the same as on Fig.~\ref{gev:lightcurves}. 
}
\label{gev:Xraylightc}
\end{figure}

The evolution of the hard X-ray fluxes extracted from the shock regions in Cas~A, Tycho and SN1006 (Fig.~\ref{gev:xrays}) are shown on Fig.~\ref{gev:Xraylightc}. The X-ray light curves for these SNRs demonstrate that the fluxes are almost steady. This is in agreement with the evolution of \g-rays from Cas~A and Tycho. As to SN1006, the two data points on the X-ray plot are also in favor of the constant flux while the decreasing trend in the \g-ray flux evolution  (Fig.~\ref{gev:lightcurves}) has quite low confidence. 

We have also measured the radio fluxes in Tycho SNR (from the same region around the forward shock as used for the X-rays) from the \verb+fits+ files of observations reported by \citet{2016ApJ...823L..32W}. The flux densities at 1.4 GHz were $1.37\pm1.17\un{Jy/beam}$ in 2002 and $1.38\pm1.18\un{Jy/beam}$ in 2013: the same at the beginning and at the end of the ten-years period. 

Radio evolution of Cas~A exhibits a flux decrease with the rate $0.7\mathrm{\%/yr}$  \citep{2009AJ....138..838H} which may be visible on the time-scale of 50 years. Visibility of the decreasing trend is questionable on the ten-years time interval: the flux density at 74 MHz was $19.6\pm0.7$ kJy in 1997 and $17.0\pm2.3$ kJy in 2006 \citep{2009AJ....138..838H}. 

For the sake of comparison, on Fig.~\ref{gev:Xraylightc}, we have also shown the variation in the hard X-ray flux from SN1987A which is thermal in nature \citep{2015ApJ...810..168O}. This SNR is discussed in more details in the next section. 

\begin{figure}
\resizebox{\hsize}{!}{\includegraphics{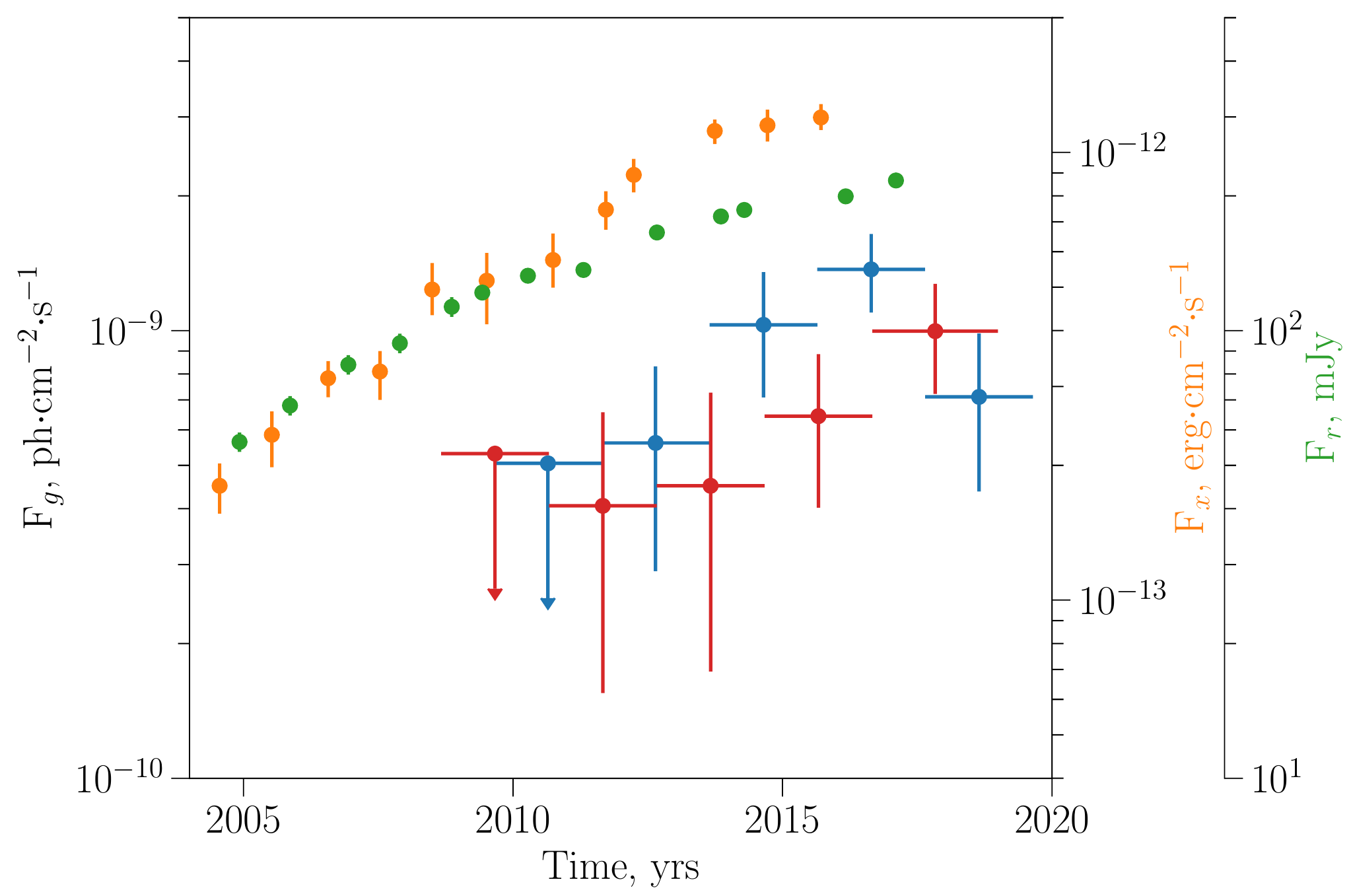}}
\caption{Radio (green), X-ray (orange) and \g-ray (blue and red) fluxes from region of SN1987A. Radio data for 8.6 GHz are from \citet{2010ApJ...710.1515Z} till 8014 day and from \citet[Table 2]{2018ApJ...867...65C} for later times. X-ray data in 3-8 keV are from \citet[][Table~1]{2016ApJ...829...40F}. Red crosses correspond to our calculations for the  \citet{2019arXiv190303045M} model; blue crosses represent our model.}
\label{gev:sn87Aall}
\end{figure}

\subsection{SN1987A}

Our results on \g-rays from SN1987A are not so solid as for Cas~A and Tycho. The main negative point is the presence of bright sources nearby \cite[Fig.~3 and 4 in][]{2016A&A...586A..71A}. The contribution of these sources to the emission in the location place of SN1987A have to be removed in order to estimate the flux from SN1987A region. Results depend critically on the models of these sources: the less one subtracts the more  could be thought as a `signal' from SNR. 

SN1987A was not previously detected in deep observations of Large Magellanic Cloud either in GeV \citep{2016A&A...586A..71A} or in TeV \citep{2015Sci...347..406H} photons. In contrast, \citet{2019arXiv190303045M} used recent data release and accurately modelled other sources in order to try to uncover emission from SN1987A.

We confirm the results of \citet{2019arXiv190303045M} though our models somehow differs. Differences consists in a shift of the observation periods, in the source catalogue (we have used 4FGL instead FL8Y), in the background emission model (we adopt  \verb+gll_iem_v07.fits+ with \verb+iso_P8R3_CLEAN_V2_v1.txt+ instead \verb+gll_iem_v06.fits+ with \verb+iso_P8R3_CLEAN_V2.txt+. It is worth noting that we have also repeated data analysis on SN1987A with (\texttt{.xml} file) sent us by D.Malyshev and with neighbouring and background sources as described in their study. With this model, we have reproduced results in their paper, including the the light curve (Fig.~\ref{gev:sn87Aall} red crosses, cf. Fig.~1 by \citet{2019arXiv190303045M}). 

Our analysis detects some flux in the region of SN1987A after the removing emission from background and nearby sources, with significance $6\sigma$, in the data collected by the Fermi observatory during 10 years till the end of August 2019. The variation of this flux in time is shown on Figs.~\ref{gev:lightcurves} and \ref{gev:sn87Aall} by the blue crosses. 

Though attribution of these \g-rays to SN1987A and even their detectability may in principle be questioned (because it depends on the assumed level of emission from the neighbouring sources and background), let us look at one feature in our results. 
Our light curve (blue on Fig.~\ref{gev:sn87Aall}) generally agrees with the trend from the  \citet{2019arXiv190303045M} model (red on Fig.~\ref{gev:sn87Aall}): the flux from the SN1987A region seems to increase with time. However, there is a difference in the last data point. In this reference, the last point is in line with brightening while in our case the flux drops. 
There is $0.28\%$ chance we may have such break randomly if the true trend consists in the brightness increase extrapolated from 2011-2017 yrs; this corresponds to $3.2\sigma$ significance, that is not strongly decisive. 
We would like to note that the period reflected by the last data point in \citet{2019arXiv190303045M} is August 2016 -- December 2018 while it is August 2017 -- August 2019 in the present paper. It seems that the difference between results is not physical. 
Low statistics remained after the elimination of the background fluxes is quite sensitive to what we actually eliminate. In order to look deeper, we compared the lists of sources involved into the analysis and found that dozens of those which has been removed from the \citet{2019arXiv190303045M} model because of the negative TS, have positive TS and remained in our model; and vice versa.

The slope of the \g-ray flux evolution generally agrees with the X-ray and radio light curves (Fig.~\ref{gev:sn87Aall}). It should be noted that X-rays from SN1987A are thermal in nature \citep{2015ApJ...810..168O} and increase in the flux after $2011$ yr is related to 
the influence of the equatorial ring \citep{2015ApJ...810..168O}. Interestingly, that such manifestation is absent in the radio data 
\citep{2019A&A...622A..73O}: green data points on our Fig.~\ref{gev:sn87Aall} do not follow the orange data points \citep[see also Fig.~10 in][]{2010ApJ...710.1515Z}.

\subsection{SN1006}

Situation with the GeV emission from SN1006 is even less confident than in the case of SN1987A. Our estimate of the signal significance is $4.8\sigma$ while typically accepted value for `detection' is at least $5\sigma$. 
This is similar to those in the previous publications. This SNR was found with the detection significance $4.7\sigma$ by \citet{2016ApJ...823...44X}. An updated analysis by the same authors stated $4.5\sigma$ for NE and $4.8\sigma$ for SW parts of the remnant \citet{2019PASJ...71...77X}. 
If one plays a bit with a source models then one can find a model which give better TS: \citet{2017ApJ...851..100C} reported significance between $5.3\sigma$ and $5.9\sigma$ depending on the source model. We have used standard models for all SNRs which we analysed in the present paper, namely, the source models from the 4FGL catalogue. 

The ratio of surface brightness between different regions of SN1006 map in different photon energy bands may shed light on some properties of SNR \citep{2012MNRAS.419..608P}. In particular, under assumption that \g-ray emission from SN1006 is the inverse-Compton process, the ratios of the brightness between the two limbs in the radio ${\cal R}\rs{r}$ and in GeV \g-rays ${\cal R}\rs{\gamma}$ may be used to see whether magnetic field strengths $B$ are similar in the these limbs:
\begin{equation}
    {\cal R}\rs{B}=\left({\cal R}\rs{r}/{\cal R}\rs{\gamma}\right)^{2/(s+1)}
\end{equation}
where ${\cal R}\rs{B}=B\rs{NE}/B\rs{SW}$ and $s$ is the radio spectral index.
The ratio is ${\cal R}\rs{r}=1.0$ in radio \citep{2012MNRAS.419..608P}.
Table~\ref{gev:table-iii} shows the GeV fluxes from NE and SW parts of SN1006 as estimated in two publications. 
Their ratio (NE to SW) is ${\cal R}\rs{\gamma}=7.0\pm3.5$ \citep{2017ApJ...851..100C} or 
${\cal R}\rs{\gamma}=0.80\pm0.53$ \citep{2019PASJ...71...77X}. The two results are quite different and we cannot draw any conclusion unless that, in the second result, the fluxes from the two regions have similar significance and they are compatible with the same magnetic field strength in both limbs.

\begin{table}
\caption{GeV fluxes from the two parts of SN006.}
\centering
\begin{tabular}{lllll}
\hline\hline
SNR  &  energy,  & TS & flux & units \\
part &  GeV &    &      &  \\
\hline 
NE$^{(1)}$& $1-2000$&28 &$6.14\pm2.53$ &$10^{-12}\un{ erg\ cm^{-2}s^{-1}}$\\
SW$^{(1)}$& &13 &$0.88\pm0.24$ &\\
\hline 
NE$^{(2)}$&$0.5-500$ &20   &$1.6\pm0.7$ &$10^{-10}\un{ ph\ cm^{-2}s^{-1}}$\\
SW$^{(2)}$& &23   &$2.0\pm1.0$&\\
\hline
\hline
\end{tabular}
\tablebib{(1)~\citet{2017ApJ...851..100C}; (2)~\citet{2019PASJ...71...77X}}
\label{gev:table-iii}
\end{table}

\begin{acknowledgements}
We are thankful to Denys Malyshev for giving us their \texttt{.xml} file for analysis of Fermi LAT observations of the region SN1987A; we used it for cross-check of the results. We thank Laura Chomiuk for providing us with the \texttt{.fits} files of the radio observations of Tycho SNRs. VM was supported by Polish NSC grant 2016/22/E/ST9/00061. MP acknowledges support of Astronomical Observatory of the Jagiellonian University of her International Summer Student Internship, where part of this work was done. This work was supported by the Ukrainian NAS project 0118U004941. 
\end{acknowledgements}

   \bibliographystyle{aa} 
   \bibliography{gevsnrs} 

\end{document}